\documentclass[prl,aps,twocolumn,superscriptaddress]{revtex4}

\nonstopmode

\usepackage{times}
\usepackage{amsmath}
\usepackage{graphicx}
\usepackage{amssymb}
\usepackage{color}

\newcommand{\p}{\partial}
\newcommand{\e}{\mathrm{e}}

\renewcommand{\Re}{\mathop{\mathrm{Re}}}
\renewcommand{\Im}{\mathop{\mathrm{Im}}}
\providecommand{\Z}{\mathbb{Z}}
\renewcommand{\Z}{\mathbb{Z}}

\providecommand{\R}{\mathbb{R}}

\begin{document}

\title{Stability of solitary waves and vortices in a 2D nonlinear Dirac model}

\author{Jes\'us Cuevas--Maraver}
\affiliation{Grupo de F\'{\i}sica No Lineal, Departamento de F\'{\i}sica Aplicada I, Universidad de Sevilla. Escuela Polit{\'e}cnica Superior, C/ Virgen de \'Africa, 7, 41011-Sevilla, Spain\\
Instituto de Matem\'aticas de la Universidad de Sevilla (IMUS). Edificio Celestino Mutis. Avda. Reina Mercedes s/n, 41012-Sevilla, Spain}

\author{Panayotis G. Kevrekidis \thanks{%
Email: kevrekid@math.umass.edu}}
\affiliation{Department of Mathematics and Statistics, University of Massachusetts,
Amherst, MA 01003-4515, USA}

\affiliation{Center for Nonlinear Studies and Theoretical Division, Los Alamos National Laboratory, Los Alamos, New Mexico 87545, USA}

\author{Avadh Saxena}
\affiliation{Center for Nonlinear Studies and Theoretical Division, Los Alamos National Laboratory, Los Alamos, New Mexico 87545, USA}

\author{Andrew Comech}
\affiliation{Department of Mathematics, Texas A\&M University, College Station, TX 77843-3368}
\affiliation{Institute for Information Transmission Problems, Moscow 127994, Russia}

\author{Ruomeng Lan}
\affiliation{Department of Mathematics, Texas A\&M University, College Station, TX 77843-3368}

\date{\today}

\begin{abstract}
We explore a prototypical {\it two-dimensional} {\it massive}
model of the nonlinear
Dirac type
and examine its solitary wave and vortex solutions. In addition
to identifying the stationary states, we provide a systematic spectral
stability analysis, illustrating the potential of spinor solutions
to be neutrally
stable in a wide parametric interval of frequencies.
Solutions of higher vorticity
are generically unstable and split into
lower charge vortices in a way that preserves
the total vorticity.
These conclusions are found not to be restricted to the case
of cubic two-dimensional nonlinearities but are found to be extended
to the case of quintic nonlinearity, as well as to that
of three spatial dimensions.
Our results also reveal
nontrivial differences with respect to
the better understood non-relativistic analogue of the model, namely
the nonlinear Schr{\"o}dinger equation.
\end{abstract}
\maketitle

{\it Introduction.} In the context of dispersive nonlinear wave
equations, admittedly the prototypical model that has attracted a wide
range of attention in optics, atomic physics, fluid mechanics,
condensed matter and mathematical physics is
the nonlinear Schr{\"o}dinger equation
(NLS)~\cite{kivshar,becbook1,becbook2,
rcg:BEC_book2,ablowitz1,sulem,chap01:bourgain}.
By comparison, far less attention has been paid to
its relativistic analogue, the nonlinear Dirac equation (NLD)~\cite{nogami},
despite its presence for almost 80 years in the context of high-energy
physics~\cite{jetp.8.260,PhysRev.83.326,RevModPhys.29.269,thirring,gavri}.
This trend is slowly starting to change, arguably,
for three principal reasons. Firstly, significant steps have been taken
in the nonlinear analysis of stability of such models~\cite{boucuc,DEP1,DEP2,DEP3,arXiv:1407.0606,linear-a},
especially in the one-dimensional (1d) setting.
Secondly, computational advances have enabled a better understanding
of the associated solutions and their
dynamics~\cite{sihong_rev,cooper,niur_recent,MR2892774,avadhj}.
Thirdly, and perhaps most importantly,
NLD starts emerging in physical systems
which arise in a diverse set of contexts of considerable interest.
These contexts include, in particular,
bosonic evolution in honeycomb lattices~\cite{Carr1,Carr2} and
a growing
class of atomically thin 2d Dirac materials~\cite{diracmat} such as graphene,
silicene, germanene and transition metal dichalcogenides \cite{tmdc}.
Recently, the physical aspects of nonlinear optics,
such as light propagation in
honeycomb photorefractive lattices
(the so-called photonic graphene)~\cite{ablowitz3,ablowitz4}
have prompted the consideration of intriguing dynamical
features, e.g. conical
diffraction in 2d honeycomb lattices~\cite{peleg}.
Inclusion of nonlinearity is then quite natural in these models, although
in a number of them (e.g., in atomic and optical physics)
the nonlinearity does not couple the spinor components.

These physical aspects
have also led to a discussion of potential 2d solutions of NLD
in~\cite{Carr1,Carr2}.
However, a systematic and
definitive characterization of stability and
nonlinear dynamical evolution
of the prototypical coherent structures in NLD models
is still lacking, to the best of our knowledge.
The present work is dedicated to offering analytical and numerical
insights into these crucial mathematical
and physical aspects of higher-dimensional
nonlinear Dirac equations
bearing in mind
the physical relevance and potential observability of such
waveforms.
As our model of choice, in order
to also be able to compare and contrast with the multitude of existing
1d results (e.g. \cite{arXiv:1407.0606,MR2892774}),
we select the well-established
Soler model~\cite{Soler}
(known in 1d as the Gross--Neveu model~\cite{Gross1}),
which is a Dirac equation with scalar self-interaction.
Such self-interaction
is based on including
into the Lagrangian density
a function of the quantity $\bar\psi\psi$
(which transforms as a scalar
under the Lorentz transformations):
\begin{equation}\label{nld-lagrangian}
\mathcal{L}\sb{\mathrm{Soler}}
=
\bar\psi
\left(i\gamma\sp\mu\partial\sb\mu-m \right) \psi
+\frac{g}{2}\left(\bar\psi\psi\right)^2,
\end{equation}
where
$m>0$, $g>0$,
$\psi(x,t)\in\mathbb{C}^N$,
$x\in\mathbb{R}^n$
and $\gamma\sp\mu$,
$0\le\mu\le n$
are $N\times N$ Dirac $\gamma$-matrices
satisfying the anticommutation relations
$\{\gamma\sp\mu,\gamma\sp\nu\}=2\eta\sp{\mu\nu}$,
with $\eta\sp{\mu\nu}$ the Minkowski tensor~\cite{dewitt},
and $\bar\psi=\psi\sp\ast\gamma\sp 0$.
(The Clifford Algebra theory gives the relation
$N\ge 2^{[(n+1)/2]}$
between the spatial dimension and spinor components
\cite[Chapter 1, \S5.3]{MR1401125}.)
The
nonlinearity of the model is generalized in the spirit of~\cite{cooper},
by using $g (\bar\psi\psi)^{k+1}/(k+1)$
with $k>0$.
The proof of existence of solitary waves in this model (in 3d)
is in \cite{MR847126,MR949625,MR1344729}.

Our results
show that the NLD in 2d admits different solutions
involving a structure of vorticity $S\in\mathbb{Z}$
in the first spinor component, with the other spinor component
bearing a vorticity $S+1$.
We identify such solutions for $S=0,\,1,\,\dots\,\,$.
While prior stability results
have often been inconclusive
(particularly in higher dimensions, see, e.g.,~\cite{PhysRevLett.50.1230}),
our
numerical computation of the spectrum
of the corresponding
linearization operator reveals that {\it only} the $S=0$ solutions
can be spectrally stable (the spectrum of the linearization
contains no eigenvalues with positive real part),
and that this stability takes place
in a rather wide interval of the frequency of the solitary waves.
On the contrary, we find that the states of higher vorticity
are generically linearly unstable.
Complementing the stability analysis results,
our direct dynamical evolution studies
show that
the unstable higher vorticity solutions
break up into lower vorticity waveforms, yet
conserving the total vorticity.
Importantly, the fundamental $S=0$ solutions are found to be potentially stable
in models both with a higher order (quintic) two-dimensional
nonlinearity, as well as
in higher dimensions (3d) under cubic nonlinearity.
These features again reflect differences
from the NLS model and as such suggest the particular interest towards
a broader and deeper study of NLD models.

An important extension of our stability findings
for higher dimensional $S=0$ solutions,
is that they remain valid for other types of nonlinearities.
These include non-Lorentz-invariant ones such as
most notably
those arising in atomic~\cite{Carr1,Carr2}, and optical~\cite{ablowitz3,ablowitz4} problems.
The fundamental difference of those models is that they correspond to
{\em massless} equations, contrary to the Soler model. For this reason, we have confirmed our stability conclusions by comparison with those emerging from the model for square binary waveguides
\cite{Biancalana} which leads to a {\em massive} nonlinear Dirac equation with the same nonlinearity as in~\cite{Carr1,Carr2}.

{\it Theoretical Setup.}
We start from the prototypical
2d nonlinear Dirac equation system,
derived from the Lagrangian density (\ref{nld-lagrangian})
with $k=1$ and $m=g=1$:
\begin{eqnarray}\label{eq:dyn}
    i\partial_t\psi_1 &=& -(i\partial_x+\partial_y)\psi_2+f(\bar\psi\psi)\psi_1 ,\nonumber \\
    i\partial_t\psi_2 &=& -(i\partial_x-\partial_y)\psi_1-f(\bar\psi\psi)\psi_2 ,
\end{eqnarray}
where $\psi_1$, $\psi_2$
are the components of the spinor $\psi\in\mathbb{C}^2$ and
the nonlinearity is
{$f(\bar\psi\psi)
=1-(\bar\psi\psi)^k
=1-(|\psi_1|^2-|\psi_2|^2)^k$.}
We note that
\eqref{eq:dyn}
is a $\mathbf{U}(1)$-invariant,
translation-invariant Hamiltonian system.

We simplify our analysis
by using the polar coordinates, where Eq.~(\ref{eq:dyn}) takes the form
\begin{eqnarray}\label{eq:stat}
    i\partial_t\psi_1
&=& -\e^{-i\theta}\left(i\partial_r+\frac{\partial_\theta}{r}\right)\psi_2+f(\psi_1,\psi_2)\psi_1,
\nonumber \\
    i\partial_t\psi_2
&=& -\e^{i\theta}\left(i\partial_r-\frac{\partial_\theta}{r}\right)\psi_1-f(\psi_1,\psi_2)\psi_2.
\end{eqnarray}
The form of this equation suggests that we look for solutions
as $\psi(\vec r,t)=\exp(-i\omega t)\phi(\vec r)$ with
\begin{equation}
    \phi(\vec r)=\left[\begin{array}{c}
    v(r)\e^{i S\theta} \\
    i\,u(r)\e^{i(S+1)\theta}
    \end{array}\right] ,
\label{ansatz}
\end{equation}
with $v(r)$ and $u(r)$ real-valued.
The value $S\in\mathbb{Z}$ can be cast as the vorticity of the first spinor component.

Once solitary waves have been identified, we explore
their stability. This approach has been
previously developed in related settings including the
multi-component NLS (see
e.g.~\cite{mussli}), as well as a massless variant of the
Dirac equation of~\cite{hc_rlse}. The presence of a mass in our
case allows not only a direct comparison with NLS (when
$\omega \rightarrow m \equiv 1$), but also generates fundamental
differences between our results and those of~\cite{Carr1,Carr2,hc_rlse},
as discussed below as well.

To examine its spectral stability, we consider a solution $\psi$
in the form of a perturbed solitary wave solution:
\begin{equation}
\psi(\vec r,t)
=\left[
\begin{array}{c}
(v(r)+\rho_1(r,\theta,t))\e^{i S\theta}
\\
i\big(u(r)+\rho_2(r,\theta,t)\big)\e^{i (S+1)\theta}
\end{array}
\right]
\e^{-i\omega t},
\end{equation}
with $\rho=(\rho_1,\rho_2)^T\in\mathbb{C}^2$ a small perturbation.
We consider the linearized equation
on $\rho$,
\begin{equation}\label{p-x-a-x}
\p\sb t
R
=
A\sb\omega
R,
\end{equation}
with
$R(r,\theta,t)
=\left(\Re\rho,\Im\rho\right)^T\in\mathbb{R}^4$
and with
a matrix-valued first order differential operator
$A\sb\omega(r,\theta,\p\sb r,\p\sb\theta)$
\cite{Supp}.
If the spectrum of the linearization operator $A\sb\omega$
contains an eigenvalue $\lambda\in\sigma(A\sb\omega)$
with $\Re\lambda>0$,
we say that the solitary wave is linearly unstable;
in such cases,
we resort to dynamical simulations of
Eqs.~(\ref{eq:dyn})
to explore the outcome of the unstable evolution.
If there are no such eigenvalues,
the solitary wave is called spectrally stable.

A convenient feature of NLS ground states
is that the linearization operator at such states,
albeit non-selfadjoint,
has its point spectrum confined
to the real and imaginary axes.
This observation is at the base
of the VK criterion \cite{VaKo}:
a linear instability can thus develop
when a positive eigenvalue bifurcates from $\lambda=0$.
More precisely,
the loss of stability due to the appearance of a pair
of positive and a pair of negative eigenvalues
follows the jump in size of the Jordan block
corresponding to the unitary invariance;
this happens when the VK condition
$\p\sb\omega Q(\omega)=0$ is satisfied,
with $Q(\omega)$ being the charge of a solitary wave.

Crucially, in the NLD case,
the spectrum of the linearization at a solitary wave
is no longer confined to the real and imaginary axes;
the linear stability analysis requires that
one studies the {\it whole complex plane}.
The key observation is that $A\sb\omega$ in \eqref{p-x-a-x}
contains $r$, $\p\sb r$, $\p\sb\theta$, but not $\theta$;
this allows to perform a detailed study of the spectrum
of $A\sb\omega$
using the decomposition of spinors into
Fourier harmonics corresponding to
different $q\in\Z$ \cite{Supp}.

In the 3d case, we are not yet able to perform
the general spectral analysis, but we studied the part of spectrum
in the invariant subspace corresponding to
perturbations of the same angular structure as
the solitary waves \cite{wakano-1966},
$[v(r)[1,0],iu(r)[\cos\theta,e^{i\phi}\sin\theta]]^T$;
this invariant subspace seems most important
since it is responsible for the linear instability in the
non-relativistic limit $\omega\to 1$
which is a consequence of the instability of the 3d cubic NLS.

\begin{figure}
\begin{tabular}{cc}
\includegraphics[width=4cm]{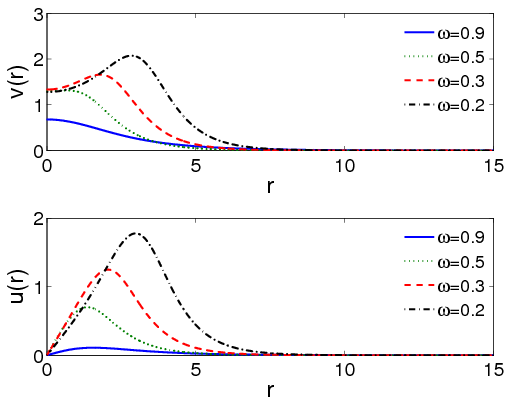} &
\includegraphics[width=4cm]{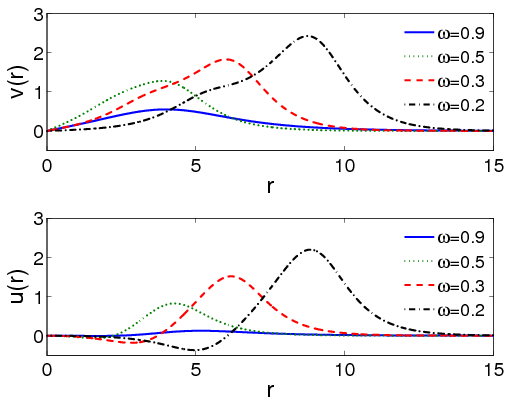} \\
\end{tabular}
\caption{Radial profiles of the spinor components for (left) $S=0$ solitary waves and (right) $S=1$ vortices for different values of $\omega$.}
\label{fig:profile}
\end{figure}

{\it Numerical results.}
We have analyzed the existence and stability of solitary waves
($S=0$, with its first component radially symmetric
and the second component having vorticity $1$)
and vortex solutions
($S=1$, with its components having vortices of order
one and two, respectively).
Both solitary waves
and vortex solutions exist in the frequency interval $\omega\in(0,m=1)$, a feature critically distinguishing our
models from those of~\cite{Carr1,Carr2}.
An intriguing
feature of the relevant waveforms is that
both the radial profile of the solitary waves and that of the vortices
possess a maximum that shifts from $r=0$ to a larger $r$ when $\omega$ approaches zero
(see Fig. \ref{fig:profile}), in a way reminiscent of the corresponding
1d solitary wave structures~\cite{cooper}.
Here the relevant state will feature a stationary bright intensity ring.
In order to obtain and analyze such coherent structures, we have made use of
the numerical methods detailed in \cite{Supp}.
To confirm the results,
we also computed the spectra
using the Evans function approach of~\cite{MR2892774}
adapted to the {present problem}.

\begin{figure}
\begin{tabular}{cc}
\includegraphics[width=4cm]{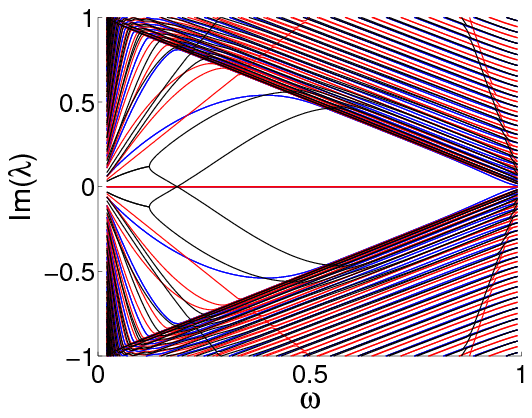} &
\includegraphics[width=4cm]{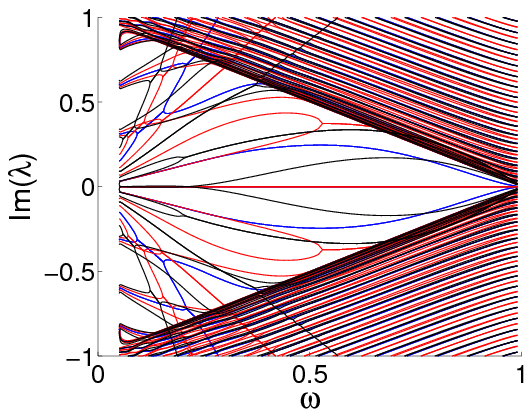} \\
\includegraphics[width=4cm]{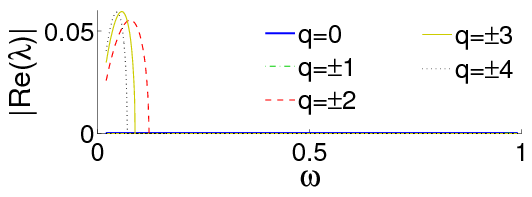} &
\includegraphics[width=4cm]{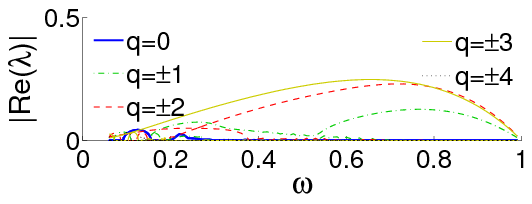} \\
\end{tabular}
\caption{
2d Soler model with cubic ($k=1$) nonlinearity.
Dependence of the (top) imaginary and (bottom) real part of the eigenvalues with respect to $\omega$.
Left (respectively, right)
panels correspond to $S=0$ solitary waves ($S=1$ vortices).
For the sake of clarity,
we only included the values
{$|q|\le 2$ for the imaginary part and $|q|\le 4$ for the real part. In the former case, the imaginary part of the eigenvalues for $q=0$, $q=\pm1$ and $q=\pm2$ are represented by, respectively, blue, red and black lines.}}
\label{fig:spectra}
\end{figure}

We start by considering the stability of $S=0$ solitary waves in the cubic ($k=1$) case. Figure \ref{fig:spectra} shows the dependence of the real and imaginary parts of the eigenvalues with respect to the stationary solution frequency
$\omega$. From the spectral dependencies we can deduce several features of
the 2d NLD equation:
(1)
It is known that the 2d NLS equation is charge-critical, and the zero eigenvalues
are degenerate~\cite{sulem}: they have higher algebraic multiplicity.
In the NLD case, however, this degeneracy is resolved: in the $S=0$ case,
as $\omega$ starts decreasing,
two eigenvalues (corresponding to $q=0$)
start at the origin when $\omega=1$
and move out of the origin for $\omega\lesssim 1$.
The absence of the algebraic degeneracy of the zero eigenvalue
{\it prevents solitary waves from NLS-like self-similar blow-up}
which is possible in charge-critical NLS \cite{MR1048692}.
(2)
The $\mathbf{U}(1)$ symmetry and the translation symmetry of the model
result in zero eigenvalues
with $q=0$ and $|q|=1$, respectively
(in both $S=0$ and $S=1$ cases).
(3)
As in the 1d NLD equation,
there are also the eigenvalues $\lambda=\pm 2\omega i$
which are associated with the $\mathbf{SU}(1,1)$
symmetry of the model~\cite{ptnlde}.
This eigenvalue pair corresponds to $q=-(2S+1)$, i.e., to
a highly excited linearization eigenstate.
(4)
Contrary to the 1d case, where the solitary waves
corresponding to any $\omega<1$
are spectrally stable,
the $S=0$ solitary wave is linearly unstable for $\omega<0.121$
because of the emergence
of nonzero real part eigenvalues
via a {\it Hamiltonian Hopf bifurcation}
in the $|q|=2$ spectrum
at $\omega=0.121$.
Another Hopf bifurcation occurs
corresponding to $|q|=3$
(at $\omega=0.0885$),
then yet another one corresponding to $|q|=4$.

\begin{figure}
\begin{tabular}{cc}
\includegraphics[width=4cm]{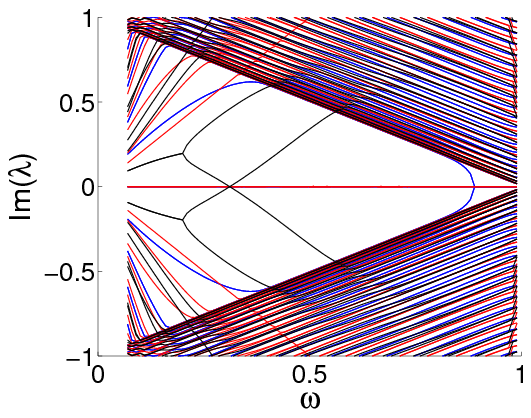} &
\includegraphics[width=4cm]{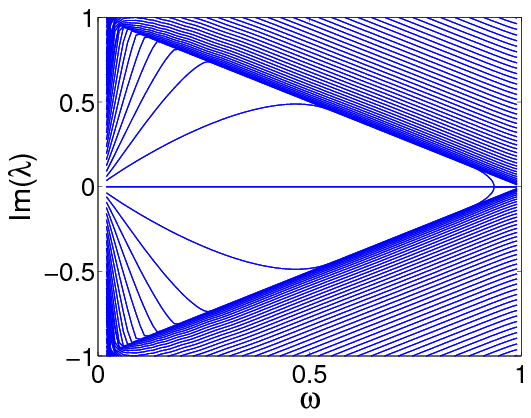} \\
\includegraphics[width=4cm]{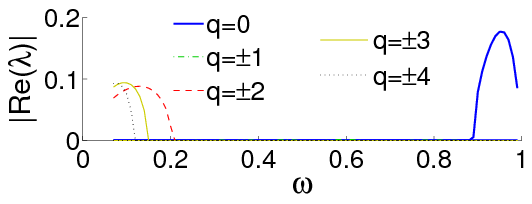} &
\includegraphics[width=4cm]{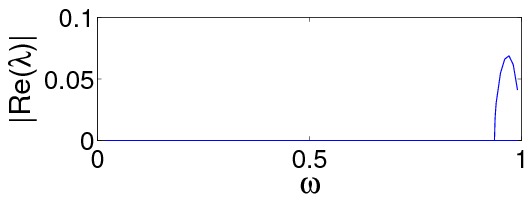} \\
\end{tabular}
\caption{
Left:
Solitary waves in the 2d Soler model with quintic ($k=2$) nonlinearity. Dependence of the (top) imaginary and (bottom) real part of the eigenvalues with respect to $\omega$ in the same format as the previous figure. Right: Solitary waves in the 3d Soler model with cubic ($k=1$) nonlinearity. Spectrum of the linearization in the one-dimensional invariant $(q=0)$ subspace which contains the eigenvalue that is responsible for the instability for $\omega\in(\omega_c,1)$, with $\omega_c\approx 0.936$.}
\label{fig:spectra-n2k2-n3k1}
\end{figure}

It is especially interesting that a
wide parametric (over frequencies) interval of
stability of solitary waves can {\it also} be observed
in the quintic ($k=2$) NLD case (see Fig.~\ref{fig:spectra-n2k2-n3k1});
while the quintic NLS solitary waves blow up (even in one
dimension), the quintic NLD solitary waves are stable even in two
dimensions, except for the interval $\omega<0.312$ where the coherent
structures experience the same Hopf bifurcation as in the cubic case, and
for $\omega>0.890$ where an exponential instability created by radial $q=0$ perturbations emerges.
Perhaps even more remarkably, the right panel
of the Fig.~\ref{fig:spectra-n2k2-n3k1} illustrates
that this stability of
NLD solitons against radial perturbations can be found in
suitable frequency intervals {\it even in 3d} (see \cite{CGG14} for a
discussion of the equations for existence and stability of radial perturbations in 3d).
Both of the above cases (quintic 2d and cubic 3d NLD)
are charge-supercritical
i.e., the charge goes to infinity in the nonrelativistic limit
$\omega\to m$.
Contrary to the pure-power supercritical NLS
whose solitary waves remain linearly unstable
for all frequencies,
solitary waves in the Soler model become spectrally stable
when $\omega$ drops below some dimension-dependent critical value
$\omega_c=\omega_c(n,k)$ \cite{Supp}.
The relevant unstable eigenvalue
(associated with $q=0$ and radially-symmetric collapse)
is only present as real
for $\omega\in(\omega_c,1)$,
where $\omega_c\approx 0.936$.
This was identified in \cite{Soler}
as the value at which both the energy and charge
of solitary waves have a minimum. Hence, we indeed find that
the radially-symmetric collapse-related
instability ceases to be present below this critical point.
Finally, as regards two dimensions,
$S=1$ vortices are unstable for every $\omega$,
because of the presence in the spectrum
of quadruplets of complex eigenvalues.
These quadruplets emerge (and disappear) for different values
of $q$
via direct (inverse Hopf) bifurcations; see the right panel
of Fig.~\ref{fig:spectra}.
The spectrum for $S=2$ vortex is quite similar to that of $S=1$;
for this reason, we do not analyze it further.

\begin{figure}
\begin{center}
\includegraphics[width=7cm,clip=on]{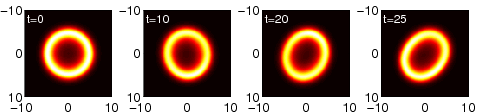}
\includegraphics[width=7cm,clip=on]{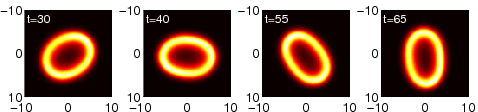}
\end{center}
\caption{Snapshots showing the evolution of the density of an
unstable $S=0$ solitary wave with $\omega=0.12$. The soliton which initially had a circular shape becomes elliptical and rotates around the center of the original solitary wave.}
\label{fig:dynS0}
\end{figure}

\begin{figure}
\includegraphics[width=7cm,clip=on]{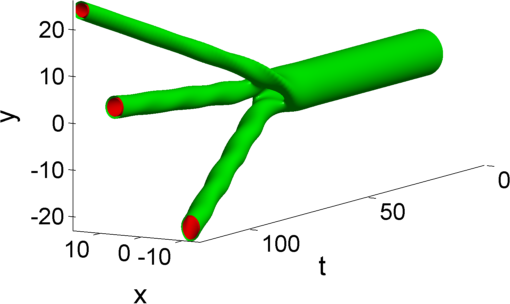}
\includegraphics[width=7cm,clip=on]{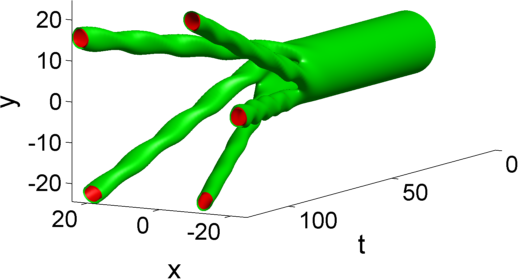} 
\includegraphics[width=7cm,clip=on]{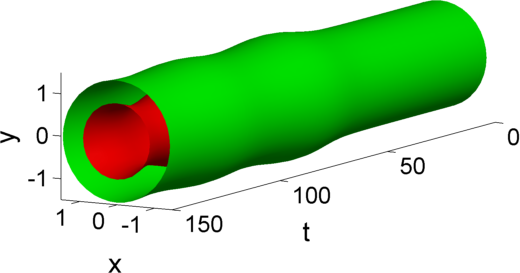}
\caption{Isosurfaces for the density of an $S=1$ (top) and $S=2$ (center) vortex with $k=1$, $\omega=0.6$ and (bottom) an $S=0$ solitary wave with $k=2$ and $\omega=0.94$.}
\label{fig:dynS12}
\end{figure}

In order to analyze the result of instabilities
{in 2d settings},
we have
probed the dynamics of unstable solutions directly (see \cite{Supp} for details).
Prototypical examples of unstable $S=0$ solitary waves and $S=1$ vortices for $k=1$ are shown in Figs. \ref{fig:dynS0} and \ref{fig:dynS12}. As
can be observed, the $S=0$ solitary waves spontaneously amplify perturbations breaking the
radial symmetry in their density and, as a result, become elliptical and rotate around the center of the circular density
of the original solitary wave
in line with the expected amplification
of the $q=2$ unstable eigenmode.
On the other hand, the $S=1$ vortices split into three smaller ones. Let us mention that in the latter case,
the first spinor component splits into
structures without angular dependence,
whereas the second component splits into
corresponding ones with angular dependence
$\propto e^{i \theta}$, in accordance with the
ansatz of Eq.~(\ref{ansatz}).
This preserves the total vorticity
across the two components, as is also shown in Fig.~\ref{fig:dynS12}. Along a similar vein, the instability of
an $S=2$ vortex eventually leads to the emergence of five
$(0,1)$ pairs, again preserving the total vorticity.
{Finally, we have analyzed the outcome of the instabilities caused by
radially-symmetric perturbations in the $k=2$ case for $\omega>\omega_c$ (see Fig. \ref{fig:dynS12}). We can observe the typical behavior of such solutions, i.e. the density width (and
amplitude) oscillate leading to a ``breathing'' structure,
but there is no collapse. This phenomenology is
reminiscent of the 1d case~\cite{nldebook}.}

{\it Conclusions and Future Challenges.}
We
have illustrated that
solitary waves of vorticity
$S=0$ in one spinor component and $S=1$ in the other are
{\it spectrally stable within a  large parametric interval}, suggesting their
physical relevance.
In that connection,
we highlight that although our
models of choice may bear a particular nonlinearity, our results
suggest that under different nonlinearities including the more
physically relevant ones of e.g.,~\cite{Carr1,Carr2} and massive
models~\cite{Biancalana}
{\it still} bear stable solitary waves for a suitable {\it wide
parametric range} of
frequencies. Thus, the conclusion of higher dimensional stability
is more general than the specifics of our particular nonlinearity
and hence of broad interest.
We also showcased the significant difference of NLD
from the
focusing  NLS equation, where
solitary waves are linearly unstable in the
charge-supercritical cases.
When the NLD solutions were found to be unstable,
their
dynamical evolution suggested breathing
oscillations in the $S=0$ case
and splitting into lower
charge configurations for $S=1$ and $S=2$.

It is of interest
to extend present considerations to
numerous settings. From a mathematical
physics perspective,
it would be useful
to explore further the 3d stability and associated dynamics.
This is  especially timely given that the 3d analogue of photonic
graphene has been experimentally realized very recently~\cite{soljacic}.
Admittedly, the latter setting does not feature a mass in the model,
thus the generalization of NLD models such as those appearing in
the works of~\cite{Carr1,Carr2,hc_rlse}
would be particularly important there.
It would
also be of interest to compare more systematically the present findings
with models associated with different nonlinearities,
including the case of honeycomb lattices in atomic and optical media
or, e.g., those stemming from wave resonances
in low-contrast photonic crystals~\cite{agueev}.

\begin{acknowledgments}
The research of A. Comech was carried out
at the Institute for Information Transmission Problems
of the Russian Academy of Sciences
at the expense of the Russian Foundation
for Sciences (project 14-50-00150). A.S. was
supported by the U. S. Department of Energy.
P.G.K. gratefully acknowledges the support of the
US-NSF under
the award
DMS-1312856, and of the
ERC under FP7, Marie Curie
Actions, People,
International Research Staff Exchange Scheme
(IRSES-605096).
\end{acknowledgments}

\newpage
\widetext

\begin{center}
\textbf{\large Stability of solitary waves and vortices in a 2D nonlinear Dirac model. Supplementary material}
\end{center}

\setcounter{equation}{0}
\setcounter{figure}{0}
\setcounter{table}{0}
\setcounter{page}{1}
\makeatletter
\renewcommand{\theequation}{S\arabic{equation}}
\renewcommand{\thefigure}{S\arabic{figure}}
\renewcommand{\bibnumfmt}[1]{[S#1]}
\renewcommand{\citenumfont}[1]{S#1}

\section{Equivalence between the Soler and massive Thirring models in 2D}

An alternative model is
the massive Thirring model (MTM) with vector self-interaction~\cite{Supp_thirring},
where the nonlinear term in the Lagrangian is based on the scalar
quantity $J\sb\mu J\sp\mu$
built from the Lorentz vector
$J\sp\mu=\bar\psi\gamma\sp\mu\psi$
which represents the charge-current density.
We mention that the MTM in one spatial dimension
is completely integrable,
and in two spatial dimensions is equivalent to the Soler model.

We demonstrate below that in two spatial dimensions the Soler and the MTM models coincide.

In the massive Thirring model with vector self-interaction, the nonlinear term in the Lagrangian is based on the scalar
quantity $J\sb\mu J\sp\mu$ built from the Lorentz vector $J\sp\mu=\bar\psi\gamma\sp\mu\psi$ which represents the charge-current density.

Using the Dirac matrices $\gamma\sp 0=\sigma_3$, $\gamma\sp j=\sigma\sb 3\sigma\sb j$, $j=1,\,2$,
and using the identity

\[(\psi\sp\ast\psi)^2=\sum\sb{j=1}\sp{3}(\psi\sp\ast\sigma_j\psi)^2\ ,\]
valid for any $\psi\in\mathbb{C}^2$, we compute that in 2d these two scalar quantities coincide:

\[
J\sb\mu J\sp\mu
=
(\psi\sp\ast\psi)^2
-(\psi\sp\ast\sigma_1\psi)^2
-(\psi\sp\ast\sigma_2\psi)^2
=(\bar\psi\psi)^2.
\]

\section{Numerical methods for the study of the existence of stationary solutions}

\subsection{Brief summary of spectral methods}

Prior to explaining the numerical methods used for calculating stationary solutions, we will proceed to present a summary of spectral methods needed for dealing with derivatives in continuum settings. For a detailed discussion on these methods, the reader is directed to \cite{Supp_Boyd} and references therein.

Spectral methods arise due to the necessity of calculating spatial derivatives with higher accuracy than that given by finite difference methods. To this aim, a differentiation matrix $\mathbf{D}\equiv\{D_{n,m}\}$ must be given together with collocation (i.e. grid) points $\mathbf{x}\equiv\{x_n\}$, which are not necessarily equi-spaced. Thus, if the {\em spectral} derivative of a function $\mathbf{f}(\mathbf{x})$ needs to be calculated, it can be cast as:

\begin{equation}
    f'(x)=\partial_x f(x) \leftrightarrow f'_n=\sum_{m=1}^N D_{n,m}f_m ,
\end{equation}
where $f_n\equiv f(x_n)$ and $f'_n\equiv f'(x_n)$. If $x\in[-L,L]$ and the boundary conditions are periodic, the Fourier collocation can be used. In this case,

\begin{equation}
    x_n=\frac{2L}{N}\left(n-\frac{N}{2}\right)\,,\quad i=1,2,\ldots N
\end{equation}
with $N$ even. The differentiation matrix is

\begin{equation}
    D_{n,m}=\left\{\begin{array}{ll} 0 & \textrm{if } n=m , \\[2ex]
    \dfrac{\pi}{2L}(-1)^{n+m}\cot\dfrac{x_n-x_m}{2} &  \textrm{if } n\neq m . \end{array}\right.
\end{equation}

Notice that doing the multiplication $\mathbf{Df}$ is equivalent to performing the following pair of Discrete Fourier Transform applications:

\begin{equation}\label{Supp_eq:FFT}
    \mathbf{Df}=\mathcal{F}^{-1}\left(i\mathbf{k}\mathcal{F}(\mathbf{f})\right)\,,
\end{equation}
with $\mathcal{F}$ and $\mathcal{F}^{-1}$ denoting, respectively, the direct and inverse discrete Fourier transform \cite{Supp_Trefethen}. The vector wavenumber $\mathbf{k}=\{k_n\}$ is defined as:

\begin{equation}
    k_n=\left\{\begin{array}{ll} \dfrac{n\pi}{L} & \textrm{if } n<N/2 , \\[2ex]
    0 &  \textrm{if } n=N/2 . \end{array}\right.
\end{equation}

The computation of the direct and inverse discrete Fourier transforms, which is useful in simulations, can be accomplished by the Fast Fourier Transform. In
what follows, however, the differentiation matrix is used for finding the Jacobian and stability matrices. Notice that the grid for finite differences discretization is the same as in the Fourier collocation; and, in addition, there is a differentiation matrix for the finite differences method, i.e.

\begin{equation}
    D_{n,m}=\frac{N}{4L}\left(\delta_{m,n+1}-\delta_{m,n-1}+\delta_{n,1}\delta_{m,N}-\delta_{n,N}\delta_{m,1}\right)\,,
\end{equation}
with $\delta$ being Kronecker's delta. It can be observed from the above
discussion that in the Fourier spectral method,
the banded differentiation matrix of the finite differences method
is substituted by a dense matrix, or, in other words,
a nearest-neighbor interaction is exchanged for into a long-range one. The
lack of sparsity of differentiation matrices is one of the drawbacks of
spectral methods, especially when having to diagonalize large systems.
However, they have the advantage of needing (a considerably) smaller number
of grid points $N$ for getting the same accuracy as with finite
difference methods.

For fixed (Dirichlet) boundary conditions, the Chebyshev spectral methods are the most suitable ones. There are several collocation schemes, the Gauss-Lobato being the most extensively used:

\begin{equation}\label{Supp_eq:gridCh}
    x_n=L\cos\left(\frac{n\pi}{N+1}\right)\,, \quad n=1,2,\ldots N\,,
\end{equation}

with $N$ being even or odd. The differentiation matrix is

\begin{equation}\label{Supp_eq:diffCh}
    D_{n,m}=\left\{\begin{array}{ll} \dfrac{x_n}{2L(1-x_n^2)} & \textrm{if } n=m , \\[2ex]
    \dfrac{(-1)^{n+m}}{L\cos(x_n-x_m)} &  \textrm{if } n\neq m . \end{array}\right.
\end{equation}

The significant drawback of Chebyshev collocation is that the discretization matrix possesses a great number of spurious eigenvalues or {\em outliers}. They are approximately equal to $N/2$. These outliers also have a significant non-zero real part, which increases when $N$ grows. This fact naturally reduces
the efficiency of the method when performing numerical time-integration. However, it presents a higher spectral accuracy than the Fourier collocation method (see e.g. \cite{Supp_Book}).

Several modifications must be introduced when applying spectral methods to polar coordinates. They basically rely on overcoming the difficulty of not having Dirichlet boundary conditions at $r=0$ and the singularity of the equations at that point. In addition, in the case of the Dirac equation, the spinor components can be either symmetric or anti-symmetric in their radial dependence, so the method described in \cite{Supp_Trefethen,Herring} must be modified accordingly. As shown in the previously mentioned references, the radial derivative of a general function $f(r,\theta)$ can be expressed as:

\begin{equation}\label{Supp_eq:polar1}
    \partial_r f(r_n,\theta)=\sum_{m=1}^N D_{n,m}f(r_m,\theta)+D_{n,2N-m}f(r_m,\theta+\pi).
\end{equation}

Notice that in this case, the collocation points must be taken as

\begin{equation}\label{Supp_eq:gridpol}
    r_n=L\cos\left(\frac{n\pi}{2N+1}\right)\,, \quad n=1,2,\ldots 2N\,,
\end{equation}
but only the first $N$ points are taken so that the domain of the radial coordinate does not include $r=0$. Analogously the differentiation matrix would
possess now $2N\times2N$ components, but only the upper half of the matrix, of size $N\times2N$ is used.

If the function that must be derived is symmetric or anti-symmetric, i.e. $f(r,\theta+\pi)=\pm f(r,\theta)$, with the upper (lower) sign corresponding to the (anti-)symmetric function, equation (\ref{Supp_eq:polar1}) can be written as:

\begin{equation}\label{Supp_eq:polar2}
    \partial_r f(r_n,\theta)=\sum_{m=1}^N \left[\left(D_{n,m}\pm D_{n,2N-m}\right)f(r_m,\theta)\right]
\end{equation}

Thus, the differentiation matrix has a different form depending whether $f(r,\theta)$ is symmetric or anti-symmetric:

\begin{equation}\label{Supp_eq:polar3}
    \partial_r\mathbf{f}(r,\theta)=\mathbf{D^{(\pm)}f}\quad \textrm{if }\ f(r,\theta)=\pm f(r,\theta+\pi)
\end{equation}
with $\mathbf{D^{(\pm)}f}$ defined as in (\ref{Supp_eq:polar2}).

\subsection{Existence of stationary solutions}

Having established the basic features of spectral methods, we have enough tools for undertaking the numerical analysis of stationary states
for the Nonlinear Dirac Equation. We recall that the dynamical equations in polar coordinates read as:

\begin{eqnarray}\label{Supp_eq:dyn}
    i\partial_t\psi_1
&=& -\e^{-i\theta}\left(i\partial_r+\frac{\partial_\theta}{r}\right)\psi_2+[1-(|\psi_1|^2-|\psi_2|^2)^k]\psi_1,
\nonumber \\
    i\partial_t\psi_2
&=& -\e^{i\theta}\left(i\partial_r-\frac{\partial_\theta}{r}\right)\psi_1-[1-(|\psi_1|^2-|\psi_2|^2)^k]\psi_2.
\end{eqnarray}
and stationary solutions can be written in the form $\psi(r,\theta,t)=\exp(-i\omega t)\phi(r,\theta)$ with
\begin{equation}
    \phi(r,\theta)=
    \left(\begin{array}{c}
    \phi_1(r,\theta) \\[1ex]
    \phi_2(r,\theta)
    \end{array}\right)
    =
    \left(\begin{array}{c}
    v(r)\e^{i S\theta} \\[1ex]
    i\,u(r)\e^{i(S+1)\theta}
    \end{array}\right) ,
\end{equation}
where $v(r)$ and $u(r)$ could be chosen real-valued.

Thus, the equations for the stationary solutions read:

\begin{eqnarray}\label{Supp_eq:stat}
    \omega v &=&  \left(\partial_r+\frac{S+1}{r}\right)u+[1-(v^2-u^2)^k]v  , \nonumber \\
    \omega u &=& -\left(\partial_r-\frac{S}{r}\right) v-[1-(v^2-u^2)^k]u ,
\end{eqnarray}
with $r>0$.

There are no analytical solutions available for this system. For this reason, numerical methods must be used to this aim. Among them we have chosen to use fixed point methods, as the Newton-Raphson one \cite{Supp_numrec}, which requires the transformation of the set of two coupled ordinary differential equations (\ref{Supp_eq:stat}) into a set of $2N$ algebraic equations; this is performed by defining the set of collocation points $\mathbf{r}\equiv\{r_n\}$, and transforming the derivatives into multiplication of the differentiation matrices $\mathbf{D}^{(1)}$ and $\mathbf{D}^{(2)}$ times the vectors $\mathbf{v}\equiv\{v_n\}$ and $\mathbf{u}\equiv\{u_n\}$, respectively, being $v_n\equiv v(r_n)$ and $u_n\equiv u(r_n)$ as explained in the previous section. Thus, the discrete version of (\ref{Supp_eq:stat}) reads:
\begin{eqnarray}
    F_n^{(1)} &\equiv& (1-\omega)v_n-(v_n^2-u_n^2)^k v_n+\sum_m D_{nm}^{(2)}u_m+\frac{S+1}{r_n}u_n =0 , \nonumber \\
    F_n^{(2)} &\equiv& (1+\omega)u_n-(v_n^2-u_n^2)^k u_n+\sum_m D_{nm}^{(1)}v_m+\frac{S}{r_n}v_n=0 .
\end{eqnarray}

It is important to notice that matrices $\mathbf{D}^{(1)}$ and $\mathbf{D}^{(2)}$ correspond to either $\mathbf{D}^{(+)}$ and $\mathbf{D}^{(-)}$, depending on the symmetry of $\mathbf{v}$ and $\mathbf{u}$, which, at the same time, depends on the value of the vorticity $S$. If $S$ is even, then $\mathbf{v}$ ($\mathbf{u}$) is (anti-)symmetric and $\mathbf{D}^{(1)}=\mathbf{D}^{(+)}$ ($\mathbf{D}^{(2)}=\mathbf{D}^{(-)}$). On the contrary, if $S$ is odd, then $\mathbf{u}$ ($\mathbf{v}$) is (anti-)symmetric and $\mathbf{D}^{(1)}=\mathbf{D}^{(-)}$ ($\mathbf{D}^{(2)}=\mathbf{D}^{(+)}$).

In order to find the roots of the vector function $\mathbf{F}=(\{F^{(1)}_n\},\{F^{(1)}_n\})^T$, an analytical expression of the Jacobian matrix
\begin{equation}
    \mathbf{J}=\left(\begin{array}{cc}
    \dfrac{\partial \mathbf{F^{(1)}}}{\partial \mathbf{v}} &
    \dfrac{\partial \mathbf{F^{(1)}}}{\partial \mathbf{u}} \\[3 ex]
    \dfrac{\partial \mathbf{F^{(2)}}}{\partial \mathbf{v}} &
    \dfrac{\partial \mathbf{F^{(2)}}}{\partial \mathbf{u}}
    \end{array}\right)=
    \left(\begin{array}{cc}
    (1-\omega)-(v^2-u^2)^{k-1}[(2k+1)v^2-u^2] & 2kuv(v^2-u^2)^{k-1}+D^{(2)}+\dfrac{S+1}{r} \\[3 ex]
    -2kuv(v^2-u^2)^{k-1}+D^{(1)}-\dfrac{S}{r} & (1+\omega)-(v^2-u^2)^{k-1}[v^2-(2k+1)u^2]
    \end{array}\right)
\end{equation}
must be introduced, with the derivatives expressed by means of spectral methods and the matrix must be evaluated at the corresponding grid points. The roots of $\mathbf{F}$, $\mathbf{\phi}=(\mathbf{v},\mathbf{u})^T$, are found by successive application of $\mathbf{\phi}\rightarrow\mathbf{\phi}-\mathbf{J}^{-1}\mathbf{F}$ until convergence is attained. In our case, we have fixed as convergence condition that
$\|\mathbf{F}\|_\infty<10^{-10}$.

\section{Linear stability of stationary solutions}

To find the spectrum of linearization
at a solitary wave in two spatial dimensions,
we consider a solution $\psi$
in the form of a perturbed solitary wave solution:
\begin{equation}
\psi(\vec r,t)
=\left[
    \begin{array}{c}
    \big(v(r)+\xi_1(r,\theta,t)+i\eta_1(r,\theta,t)\big)\e^{i S\theta} \\[2ex]
\!    i\big(u(r)+\xi_2(r,\theta,t)+i\eta_2(r,\theta,t)\big)\e^{i(S+1)\theta}
    \end{array}
\!\right]
\e^{-i\omega t},
\end{equation}
with
$\xi(\vec r,t)$ and $\eta(\vec r,t)\in\R^2$
corresponding to a small perturbation.
The linearized equation
on
$R(r,\theta,t)=[\xi_1,\xi_2,\eta_1,\eta_2]^T
\in\mathbb{R}^4$
has the form
\[
\p\sb t
R
=
A\sb{\omega,S}
R,
\]
with $A\sb{\omega,S}(r,\theta,\p\sb r,\p\sb\theta)$
a matrix-valued first order differential operator
\begin{equation}\label{Supp_def-operator-a-a}
A\sb{\omega,S}(r,\theta,\p\sb r,\p\sb\theta)
=
\left[
\begin{array}{cc}
-\sigma_1\frac{\p\sb\theta}{r}
&L_{-}
\\[2ex]
-L_{+}&-\sigma_1\frac{\p\sb\theta}{r}
\end{array}
\right],
\end{equation}
where
\[
\sigma_1=\left(\begin{array}{cc}0&1\\[1ex]1&0\end{array}\right),
\qquad
L_{-}=
\left(
\begin{array}{cc}
f-\omega&\p\sb r+\frac{S+1}{r}
\\[1ex]
-(\p\sb r-\frac{S}{r})&-f-\omega
\end{array}
\right),
\qquad
L_{+}=L_{-}
+
2\left(
\begin{array}{cc}
v^2&-uv
\\[1ex]
-uv&u^2
\end{array}
\right)f',
\]
with
$v=v(r,\omega,S)$,
$u=u(r,\omega,S)$,
and
with
$f(\tau)=m-\tau^\kappa$, $f'(\tau)=-\kappa \tau^{\kappa-1}$
evaluated at
$\tau=v^2-u^2$.
To find the spectrum of the operator $A\sb{\omega,S}$,
we consider it in the space of $\mathbb{C}^4$-valued functions.
The key observation which facilitates
a computation of the spectrum
is that the explicit form \eqref{Supp_def-operator-a-a}
of $A\sb{\omega,S}$
contains $r$, $\p\sb r$, $\p\sb\theta$, but not $\theta$.
As a consequence, $A\sb{\omega,S}$
is invariant in the spaces
which correspond to the Fourier decomposition
with respect to $\theta$,
\[
X\sb q
=\left\{[a_1(r);a_2(r);b_1(r);b_2(r)]e^{iq\theta}\right\};
\]
the restriction of $A\sb{\omega,S}$
to each such subspace is given by

\begin{equation}\label{Supp_def-operator-a-a-q}
A\sb{\omega,S,q}(r,\p\sb r)
=A\sb{\omega,S}\vert_{X\sb q}
=
\left[
\begin{array}{cc}
-\sigma_1\frac{iq}{r}
&L_{-}
\\[2ex]
-L_{+}&-\sigma_1\frac{iq}{r}
\end{array}
\right],
\qquad
q\in\Z,
\end{equation}
and this allows us to compute the spectrum of $A\sb{\omega,S}$
as the union of spectra of the one-dimensional spectral problems:
\begin{equation}\label{Supp_harmonics}
\sigma\left(A\sb{\omega,S}\right)
=
\mathop{\bigcup}\limits\sb{q\in\Z}
\sigma\left(A\sb{\omega,S,q}\right).
\end{equation}

The spectrum $\sigma\left(A\sb{\omega,S,q}\right)$ is found by evaluating the functions appearing therein at the collocations points and substituting the partial derivatives by the corresponding differentiation matrix. At this point, one must be very cautious because, as also occurred with the Jacobian, there will be two different differentiation matrices in our problem. Thus, the representation of $L_-$ matrix will be:

\[
L_{-}=
\left(
\begin{array}{cc}
f-\omega & D^{(2)}+\frac{S+1}{r}
\\[2ex]
-(D^{(1)}-\frac{S}{r})&-f-\omega
\end{array}
\right)
\]

In the 3d case,
the NLD solitary waves
were initially found in \cite{Supp_wakano-1966}
to be of the form
$
\left[
\begin{array}{c}
v(r)
\left(\begin{array}{c}1\\0\end{array}\right)
\\
i u(r)
\left(\begin{array}{c}\cos\theta\\e^{i\phi}\sin\theta\end{array}\right)
\end{array}
\right]e^{-i\omega t},
$
with $\omega\in(0,1)$
and with $v$, $u$ satisfying
\begin{eqnarray}\label{Supp_eq:stat-3d}
    \omega v &=&  \left(\partial_r+\frac{2}{r}\right)u+(1-(v^2-u^2)^k)v  , \nonumber \\
    \omega u &=& -\partial_r v-(1-(v^2-u^2)^k)u ,
\qquad
r>0.
\end{eqnarray}
To study the linearization operator
in the invariant space which has the same
angular dependence as the solitary waves,
we consider the perturbed solutions in the form
\[
\psi(\vec r,t)
=
\left[
\begin{array}{c}
(v(r)+\xi_1(r,t)+i\eta_1(r,t))
\left(\begin{array}{c}1\\0\end{array}\right)
\\
i(u(r)+\xi_2(r,t)+i\eta_2(r,t))
\left(\begin{array}{c}\cos\theta\\e^{i\phi}\sin\theta\end{array}\right)
\end{array}
\right]e^{-i\omega t}.
\]
The linearized equation on
$R(r,t)=(\xi_1,\xi_2,\eta_1,\eta_2)^T$
is similar to \eqref{Supp_def-operator-a-a}:
\[
\p\sb t
R
=A\sb\omega R,
\quad
\mathrm{with}
\quad
A\sb\omega
=
\left[
\begin{array}{cc}0&L_{-}\\[2ex]-L_{+}&0\end{array}
\right],
\quad
L_{-}=
\left(
\begin{array}{cc}
f-\omega&\p\sb r+\frac{2}{r}
\\[1ex]
-\p\sb r&-f-\omega
\end{array}
\right),
\quad
L_{+}=L_{-}
+
2\left(
\begin{array}{cc}
v^2&-uv
\\[1ex]
-uv&u^2
\end{array}
\right)f'.
\]

\section{Critical frequency in 2D and 3D configurations}

The figure below shows the critical frequencies {for radially-symmetric collapse}, as a function of the exponent $k$ associated with the nonlinearity
in our generalized NLD model. For $\omega\in(\omega_c,1)$, the NLD solitary waves are linearly unstable. Below $\omega_c$ the linear instability disappears. For $k\le 2/n$, {with $n$ being the system dimension,} there is no linear instability for $\omega\lesssim 1$.

\begin{center}
\includegraphics[width=6cm]{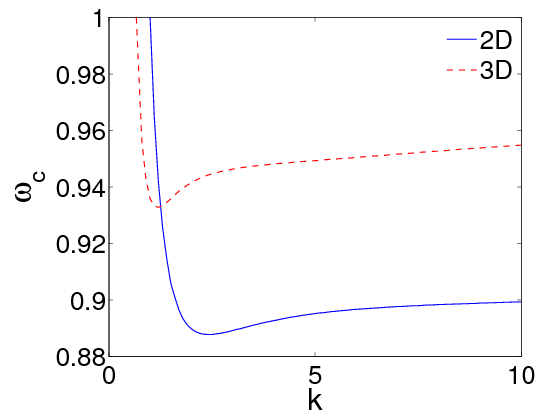}
\end{center}

\section{Dynamics}

Finally, we briefly discuss how the dynamics of Eq. (\ref{Supp_eq:dyn}) is simulated. In this case, Chebyshev spectral methods are not the most suitable ones, because of the presence of many outliers (see \cite{Supp_Book}). In addition, as it was demonstrated in \cite{Supp_jpa}, finite difference methods also raise a number
of concerns for the Dirac equation. Thus, the  possibility that appears
to us to be optimal presently is to use Fourier spectral methods, that works fairly well as long as the frequency $\omega$ is not close to zero.

Consequently, periodic boundary conditions must be supplied to our problem.
This is less straightforward when working in polar coordinates in the domain $(0,L)\times[0,2\pi)$. For this reason, we opt to work with a purely 2D problem in rectangular coordinates in the domain $(-L,L]\times(-L,L]$. The simulations of the paper have  been performed with a Dormand-Prince numerical integrator using such a spectral collocation scheme with the aid of Fast Fourier Transforms (\ref{Supp_eq:FFT}).

\end{document}